\documentclass[aps,preprint]{revtex4}%
\usepackage{amsfonts}
\usepackage{amsmath}
\usepackage{amssymb}
\usepackage{graphicx}%
\providecommand{\U}[1]{\protect\rule{.1in}{.1in}}

\begin{document}
\title{Including many-body screening into self-consistent calculations: Tight-binding
model studies with the Gutzwiller approximation}
\author{Y. X. Yao, C. Z. Wang, and K. M. Ho}
\affiliation{Ames Laboratory-U.S. DOE. and Department of Physics and Astronomy, Iowa State
University, Ames, Iowa 50011, USA}


\begin{abstract}
We introduce a scheme to include many-body screening processes explicitly into
a set of self-consistent equations for electronic structure calculations using
the Gutzwiller approximation. The method is illustrated by the application to
a tight-binding model describing the strongly correlated $\gamma$-Ce system.
With the inclusion of the $5d$-electrons into the local Gutzwiller projection
subspace, the correct input Coulomb repulsion $U_{ff}$ between the
$4f$-electrons for $\gamma$-Ce in the calculations can be pushed far beyond
the usual screened value $U_{ff}^{scr}$ and close to the bare atomic value
$U_{ff}^{bare}$. This indicates that the $d$-$f$ many-body screening is the
dominant contribution to the screening of $U_{ff}$ in this system. The method
provides a promising way towards the \textit{ab initio} Gutzwiller density
functional theory.

\end{abstract}
\maketitle

\section{Introduction}

Over the past several decades, density functional theory (DFT)\cite{DFT1} with
the local density approximation (LDA)\cite{DFT2} has been very successful in
electronic structure and total energy calculations in many systems. Meanwhile,
several hybrid approaches based on the combination of LDA with many-body
techniques have been proposed to overcome the limitation of LDA in strongly
correlated electron systems. Among these hybrid approaches,
LDA+U\cite{LDAU1,LDAU2} is the most widely used method. Using a more detailed
treatment of electronic correlation effects, LDA plus dynamical mean field
theory (LDA+DMFT)\cite{LDADMFT1,LDADMFT2} takes the local quantum fluctuations
into account and can calculate both ground state and excited state properties.
However, due to the large computational load of DMFT, LDA+Gutzwiller method
has recently been developed to calculate the ground state properties of
correlated systems\cite{LDAG1,LDAG2}. All practical calculations with the
above hybrid approaches use as input the screened Coulomb repulsion U and
Hund's coupling J parameters for the correlated orbitals, which have to be
estimated in advance using experiment data, constrained-LDA
calculations\cite{CLDA} or random phase approximation
(RPA)\cite{RPA1,RPA2,RPA3}.

Different from the above hybrid approaches which require prior determination
of the screened Coulomb repulsion U and Hund's coupling J parameters, a
Gutzwiller density functional theory (GDFT)\cite{Ho} has recently been
proposed as an \textit{ab initio} approach which directly takes the Coulomb
integrals of the local orbitals and incorporates the screening process
explicitly through a self-consistent solution of the many-electron wave
function. In the GDFT, a Gutzwiller form is adopted for the variational wave
function to go beyond single Slater determinant-based approaches, while the
effective single particle picture is retained through the Gutzwiller
approximation\cite{Gutzwiller1,Gutzwiller2,Bunemann07}.

In this paper, we address an important yet unresolved question in the GDFT,
i.e., how to choose the local subspace for the Gutzwiller operator such that
the most relevant many-body screening processes can be captured in a
self-consistent calculation. An orthogonal tight-binding model for the
strongly correlated $\gamma$-phase Ce is chosen as a prototype to examine the
effects of including main onsite screening $5d$-channels in addition to the
localized $4f$-orbitals into the Gutzwiller local subspace. In many LDA+$U$,
LDA+Gutzwiller and LDA+DMFT calculations such screening processes are usually
bypassed by introducing the screened interaction parameters (e.g., U and J
parameters) as input between the minimal correlated orbitals (e.g.,
4f-orbitals for Ce). Such screened $U$ parameter is usually much smaller than
the bare atomic value since the screening effects are assumed to be absorbed
in it. By the explicit inclusion of some many-body screening processes into
self-consistent calculations, one may expect to use the partially unscreened
larger input $U$ parameter for the calculations in order to obtain the same
physical effects. And the fully unscreened $U$ parameter (i.e., bare atomic
value) should be used as input for the calculations if all the important
many-body screening processes are explicitly dealt with. In this paper, we
show that a much larger input $U$ parameter close to bare atomic value is
needed for the calculations with both $5d$ and $4f$-orbitals in the Gutzwiller
local subspace in order to retain the same physical effects as one uses the
screened $U$ parameter with minimal $4f$-orbitals in the Gutzwiller local
subspace for $\gamma$-Ce. Therefore the treatment of $5d$-$4f$ interactions on
the Gutzwiller level is indeed a many-body screening process. Including the
many-body screening effects in a self-consistent way is critical for a
predictive first principles theory. Such calculations can also provide useful
insights on how to determine accurate screened interaction parameters.

\section{Method}

The electron Hamiltonian for $\gamma$-phase Ce is written as%
\begin{equation}
\mathcal{H}=\mathcal{H}_{0}+\mathcal{H}_{1}+\mathcal{H}_{2} \label{Hamil}%
\end{equation}
where the bare paramagnetic band Hamiltonian\ is%
\begin{equation}
\mathcal{H}_{0}=\sum_{\left(  i\alpha\right)  \neq\left(  j\beta\right)
,\sigma}t_{i\alpha j\beta}c_{i\alpha\sigma}^{\dagger}c_{j\beta\sigma}%
+\sum_{i,\alpha,\sigma}\varepsilon_{i\alpha}c_{i\alpha\sigma}^{\dagger
}c_{i\alpha\sigma}%
\end{equation}
$t_{i\alpha j\beta}$ is the electron hopping element between orbital $\alpha$
at site $i$ and orbital $\beta$ at site $j$. $\varepsilon_{i\alpha}$ is the
orbital level.$\ \alpha$ and $\beta$\ run over all the basis set orbitals,
i.e., $6s$, $5d$ and $4f$-orbitals of Ce. $c^{\dagger}$($c$) is the electron
creation (annihilation) operator. $\sigma$\ is spin index. The typical
simplified onsite term for $4f$ electrons is%
\begin{equation}
\mathcal{H}_{1}=\frac{U_{ff}}{2}\sum_{i,\gamma\gamma^{\prime}\in\left\{
4f\right\}  ,\left(  \gamma\sigma\right)  \neq\left(  \gamma^{\prime}%
\sigma^{\prime}\right)  }c_{i\gamma\sigma}^{\dagger}c_{i\gamma\sigma
}c_{i\gamma^{\prime}\sigma^{\prime}}^{\dagger}c_{i\gamma^{\prime}%
\sigma^{\prime}}%
\end{equation}
where only density-density type interactions are included. $U_{ff}$ is the
Coulomb repulsion energy between the localized $4f$-electrons. $\gamma$\ is
the orbital index in the Gutzwiller local subspace. $\left\{  4f\right\}
$\ denotes the set of $4f$-orbitals. In our present model, an additional
onsite many-body interaction term%
\begin{equation}
\mathcal{H}_{2}=U_{fd}\sum_{i,\gamma\in\left\{  4f\right\}  ,\gamma^{\prime
}\in\left\{  5d\right\}  ,\sigma\sigma^{\prime}}c_{i\gamma\sigma}^{\dagger
}c_{i\gamma\sigma}c_{i\gamma^{\prime}\sigma^{\prime}}^{\dagger}c_{i\gamma
^{\prime}\sigma^{\prime}}%
\end{equation}
is also included where $U_{fd}$ is the Coulomb repulsion energy between the
localized $4f$-electron and $5d$-electron as introduced in the Falicov-Kimball
model\cite{Falicov}. The Gutzwiller local subspace in this model includes $4f$
and $5d$-orbitals.

We use a variational wave function of the Gutzwiller form,
\begin{equation}
\left\vert \Psi_{G}\right\rangle =\frac{\hat{G}\left\vert \Psi_{0}%
\right\rangle }{\sqrt{\left\langle \Psi_{0}\left\vert \hat{G}^{2}\right\vert
\Psi_{0}\right\rangle }}%
\end{equation}
with the Gutzwiller approximation to calculate the expectation values of the
Hamiltonian \cite{Gutzwiller1,Gutzwiller2,Bunemann07}. $\Psi_{0}$ is the
uncorrelated wave function and $\hat{G}$ is the Gutzwiller projection
operator. The Gutzwiller approximation renormalizes the expectation values of
one-particle operators. The renormalization factors have analytic expressions,
however, they are usually quite complicated. These expressions can be
simplified by introducing a set of rotated orbitals $\left\{  h_{i\gamma
\sigma}^{\dagger}\right\}  $ in the Gutzwiller local subspace following
Ref.\cite{Bunemann07}, such that the local single particle density matrix is
diagonalized, i.e.,%
\begin{equation}
h_{i\gamma\sigma}^{\dagger}=\sum_{\gamma^{\prime}}u_{\gamma\gamma^{\prime}%
}^{\ast}c_{i\gamma^{\prime}\sigma}^{\dagger}%
\end{equation}%
\begin{equation}
\left\langle h_{i\gamma\sigma}^{\dagger}h_{i\gamma^{\prime}\sigma
}\right\rangle _{0}=n_{i\gamma\sigma}^{0}\delta_{\gamma\gamma^{\prime}}%
\end{equation}
Here we define $\left\langle \hat{O}\right\rangle _{0}\equiv\left\langle
\Psi_{0}\left\vert \hat{O}\right\vert \Psi_{0}\right\rangle $ for a general
operator $\hat{O}$.

We can rewrite the Hamiltonian (Eq.\ref{Hamil}) of the system in terms of the
rotated local natural basis set orbitals%
\begin{equation}
\mathcal{H}=\sum_{\left(  i\alpha\right)  \neq\left(  j\beta\right)  ,\sigma
}\tilde{t}_{i\alpha j\beta}h_{i\alpha\sigma}^{\dagger}h_{j\beta\sigma}%
+\sum_{i,\alpha,\sigma}\tilde{\varepsilon}_{i\alpha}h_{i\alpha\sigma}%
^{\dagger}h_{i\alpha\sigma}+\frac{1}{2}\sum_{i,\left(  \gamma\sigma\right)
\neq\left(  \gamma^{\prime}\sigma^{\prime}\right)  }\tilde{U}_{\gamma
\gamma^{\prime}}^{i}h_{i\gamma\sigma}^{\dagger}h_{i\gamma\sigma}%
h_{i\gamma^{\prime}\sigma^{\prime}}^{\dagger}h_{i\gamma^{\prime}\sigma
^{\prime}}%
\end{equation}
Here we define $h_{i\alpha\sigma}^{\dagger}\equiv c_{i\alpha\sigma}^{\dagger}$
for the orbitals outside the Gutzwiller local subspace and the set of
$\tilde{t}_{i\alpha j\beta}$, $\tilde{\varepsilon}_{i\alpha}$, and $\tilde
{U}_{\gamma\gamma^{\prime}}^{i}$ in the new representation can be obtained
from their values ${t}_{i\alpha j\beta}$, ${\varepsilon}_{i\alpha}$, and
${U}_{\gamma\gamma^{\prime}}^{i}$ in the original Hamiltonian through the
basis transformation.

We introduce a Gutzwiller operator in the following form%
\begin{equation}
\hat{G}=e^{-\sum_{i\mathcal{F}}g_{i\mathcal{F}}\left\vert \mathcal{F}%
_{i}\right\rangle \left\langle \mathcal{F}_{i}\right\vert }%
\end{equation}
where $\left\vert \mathcal{F}_{i}\right\rangle $ is the Fock state generated
by a set of $\left\{  h_{i\gamma\sigma}^{\dagger}\right\}  $%
\begin{equation}
\left\vert \mathcal{F}_{i}\right\rangle =%
{\displaystyle\prod\limits_{\gamma\sigma}}
\left(  h_{i\gamma\sigma}^{\dagger}\right)  ^{n_{i\gamma\sigma}^{\mathcal{F}}%
}\left\vert 0\right\rangle
\end{equation}
with $n_{i\gamma\sigma}^{\mathcal{F}}=\left\langle \mathcal{F}_{i}\left\vert
n_{i\gamma\sigma}\right\vert \mathcal{F}_{i}\right\rangle $, which identifies
whether there is an electron with spin $\sigma$ occupied in orbital $\gamma$
for Fock state $\mathcal{F}$ at the $i^{th}$\ site. $g_{i\mathcal{F}}=1$ for
empty and singly occupied configurations because in these cases there are no
electron-electron repulsion involved. According to Ref.\cite{Bunemann07}, the
expectation value of the electron Hamiltonian $\mathcal{H}$ for $\gamma$-Ce
can be expressed as%
\begin{equation}
\left\langle \mathcal{H}\right\rangle _{G}=\sum_{\alpha,\beta,\sigma
,\mathbf{k}}\left(  z_{\alpha\sigma}z_{\beta\sigma}\tilde{t}_{\alpha\beta
}^{\mathbf{k}}+\tilde{\varepsilon}_{\alpha}\delta_{\alpha\beta}\right)
\left\langle h_{\mathbf{k}\alpha\sigma}^{\dagger}h_{\mathbf{k}\beta\sigma
}\right\rangle _{0}-\sum_{\gamma,\sigma}\tilde{\varepsilon}_{\gamma}%
n_{\gamma\sigma}^{0}+\sum_{\mathcal{F}}E_{\mathcal{F}}p_{\mathcal{F}}%
\end{equation}
where $\tilde{t}_{\alpha\beta}^{\mathbf{k}}$ is the Fourier transformation
coefficient of $\tilde{t}_{i\alpha j\beta}$ at crystal momentum $\mathbf{k}$.
The site indices are dropped since there is only one atom in the primitive
unit cell of $\gamma$-Ce. We define the Gutzwiller orbital renormalization
factor $z_{\alpha\sigma}\equiv1$ for orbitals outside the Gutzwiller local
subspace. Following ref\cite{Bunemann07},%
\begin{equation}
z_{\gamma\sigma}=\frac{1}{\sqrt{n_{\gamma\sigma}^{0}\left(  1-n_{\gamma\sigma
}^{0}\right)  }}\sum_{\mathcal{F},\mathcal{F}^{\prime}}\sqrt{p_{\mathcal{F}%
}p_{\mathcal{F}^{\prime}}}\left\vert \left\langle \mathcal{F}\left\vert
h_{\gamma\sigma}^{\dagger}\right\vert \mathcal{F}^{\prime}\right\rangle
\right\vert ^{2}%
\end{equation}
for the orbitals inside the Gutzwiller local subspace. The energy of
configuration $\left\vert \mathcal{F}\right\rangle $ is%
\begin{equation}
E_{\mathcal{F}}=\left\langle \mathcal{F}\left\vert \sum_{\gamma\sigma}%
\tilde{\varepsilon}_{\gamma}h_{\gamma\sigma}^{\dagger}h_{\gamma\sigma}%
+\frac{1}{2}\sum_{\left(  \gamma\sigma\right)  \neq\left(  \gamma^{\prime
}\sigma^{\prime}\right)  }\tilde{U}_{\gamma\gamma^{\prime}}h_{\gamma\sigma
}^{\dagger}h_{\gamma\sigma}h_{\gamma^{\prime}\sigma^{\prime}}^{\dagger
}h_{\gamma^{\prime}\sigma^{\prime}}\right\vert \mathcal{F}\right\rangle
\end{equation}
Here $p_{\mathcal{F}}$ is the occupation probability of configuration
$\left\vert \mathcal{F}\right\rangle $, which satisfies the following
constraints%
\begin{equation}
\sum_{\mathcal{F}}p_{\mathcal{F}}=1 \label{C1}%
\end{equation}%
\begin{equation}
\sum_{\mathcal{F}}p_{\mathcal{F}}n_{\alpha\sigma}^{\mathcal{F}}=n_{\alpha
\sigma}^{0} \label{C2}%
\end{equation}
In practical calculations $\mathcal{H}_{0}$\ is obtained by downfolding the
first-principles LDA band structure. Therefore some contributions to the total
energy of the system from $\mathcal{H}_{1}$ and $\mathcal{H}_{2}$\ have
already been taken into account by $\mathcal{H}_{0}$\ in a mean-field way.
Such contributions are commonly referred to as the double counting term which
should be subtracted in the expression for the total energy. Therefore the
total energy per unit cell of the system is given by%
\begin{equation}
E_{T}=\left\langle \mathcal{H}\right\rangle _{G}-E_{D.C.}%
\end{equation}
Following the treatment in LDA+U calculations\cite{DC1,DC2} we choose
$E_{D.C.}$ to be:%
\begin{equation}
E_{D.C.}=\frac{1}{2}U_{ff}N_{f}\left(  N_{f}-1\right)  +U_{fd}N_{f}N_{d}
\label{dc}%
\end{equation}
where $N_{f}$($N_{d}$) is the total number of $4f$($5d$) electrons.

Minimization of the total energy $E_{T}$ with respect to the band wave
function $\psi_{n\mathbf{k}\sigma}$ and the local configuration occupation
probability $p_{\mathcal{F}}$ under the set of constraints given by
Eq.\ref{C1} and Eq.\ref{C2} yields the following set of equations which need
to be solved self-consistently,
\begin{equation}
\mathcal{H}_{eff}^{\mathbf{k}\sigma}\left\vert \psi_{n\mathbf{k}\sigma
}\right\rangle =\epsilon_{n\mathbf{k}\sigma}\left\vert \psi_{n\mathbf{k}%
\sigma}\right\rangle \label{H1}%
\end{equation}%
\begin{equation}
\sum_{\mathcal{F}^{\prime}}\mathcal{M}_{\mathcal{FF}^{\prime}}\sqrt
{p_{\mathcal{F}^{\prime}}}=\mu_{0}\sqrt{p_{\mathcal{F}}} \label{H2}%
\end{equation}
where%
\begin{equation}
\mathcal{H}_{eff}^{\mathbf{k}\sigma}=\sum_{\alpha,\beta,\sigma}\left(
z_{\alpha\sigma}z_{\beta\sigma}\tilde{t}_{\alpha\beta}^{\mathbf{k}}%
+\tilde{\varepsilon}_{\alpha}\delta_{\alpha\beta}\right)  h_{\mathbf{k}%
\alpha\sigma}^{\dagger}h_{\mathbf{k}\beta\sigma}+\sum_{\gamma,\sigma}%
\mu_{\gamma\sigma}h_{\mathbf{k}\gamma\sigma}^{\dagger}h_{\mathbf{k}%
\gamma\sigma} \label{H1.1}%
\end{equation}
and
\begin{equation}
\mathcal{M}_{\mathcal{F},\mathcal{F}^{\prime}}=\sum_{\gamma\sigma}%
\frac{e_{\gamma\sigma}}{2\sqrt{n_{\gamma\sigma}^{0}\left(  1-n_{\gamma\sigma
}^{0}\right)  }}\left\vert \left\langle \mathcal{F}\left\vert h_{\gamma\sigma
}^{\dagger}+h_{\gamma\sigma}\right\vert \mathcal{F}^{\prime}\right\rangle
\right\vert ^{2}+\delta_{\mathcal{FF}^{\prime}}\left(  E_{\mathcal{F}}%
-\sum_{\alpha\sigma}\mu_{\alpha\sigma}n_{\alpha\sigma}^{\mathcal{F}}\right)
\end{equation}
The effective single-particle Hamiltonian Eq.\ref{H1.1} has been shown to
describe the Landau-Gutzwiller quasiparticle bands\cite{Bunemann03}, and the
square of the $z$-factor corresponds to the quasiparticle weight\cite{LDAG2}.
The local orbital chemical potentials are given by
\begin{align}
\mu_{\gamma\sigma}  &  =-\tilde{\varepsilon}_{\gamma}+\frac{\partial
z_{\gamma\sigma}}{\partial n_{\gamma\sigma}^{0}}e_{\gamma\sigma}+\eta
_{\gamma\sigma}\nonumber\\
&  -\left(  U_{ff}\left(  N_{f}-\frac{1}{2}\right)  +U_{fd}N_{d}\right)
I_{\left[  \gamma\in\left\{  4f\right\}  \right]  }\nonumber\\
&  -U_{fd}N_{f}I_{\left[  \gamma\in\left\{  5d\right\}  \right]  }%
\end{align}
with%
\begin{equation}
e_{\gamma\sigma}=\sum_{\mathbf{k},\beta}\left(  z_{\beta\sigma}\tilde
{t}_{\gamma\beta}^{\mathbf{k}}\left\langle h_{\mathbf{k}\gamma\sigma}%
^{\dagger}h_{\mathbf{k}\beta\sigma}\right\rangle _{0}+c.c.\right)
\end{equation}
$\eta_{\alpha\sigma}$ is the Lagrange multiplier associated with the
constraint of Eq.\ref{C2}. $I_{\left[  x\right]  }$ is an indicator function
which equals $1$ if $x$ is true and $0$ otherwise.

To get the self-consistent solution of Eq.\ref{H1} and \ref{H2} with the
constraints of Eq.\ref{C1} and \ref{C2}, one starts with some initial guess of
$\left\{  z_{\gamma\sigma}\right\}  $ and $\left\{  \mu_{\gamma\sigma
}\right\}  $. The band wave functions $\left\{  \psi_{n\mathbf{k}\sigma
}\right\}  $ can be obtained straightforwardly by diagonalizing the effective
single electron Hamiltonian $\mathcal{H}_{eff}^{\mathbf{k}}$. Then Eq.\ref{H2}
with constraints of Eq.\ref{C1} and \ref{C2} can be solved by adjusting
$\left\{  \eta_{\gamma\sigma}\right\}  $\ such that the real symmetric matrix
$\mathcal{M}$\ generates the lowest-lying eigenvector which satisfies
Eq.\ref{C2}. $\left\{  z_{\gamma\sigma}\right\}  $ and $\left\{  \mu
_{\gamma\sigma}\right\}  $ would be updated according to the solution of
Eq.\ref{H2} until convergence is reached\cite{Thesis1}.

\section{Results and Discussions}

\begin{figure}
[ptb]
\begin{center}
\includegraphics[
height=5.7705cm,
width=8.1012cm
]%
{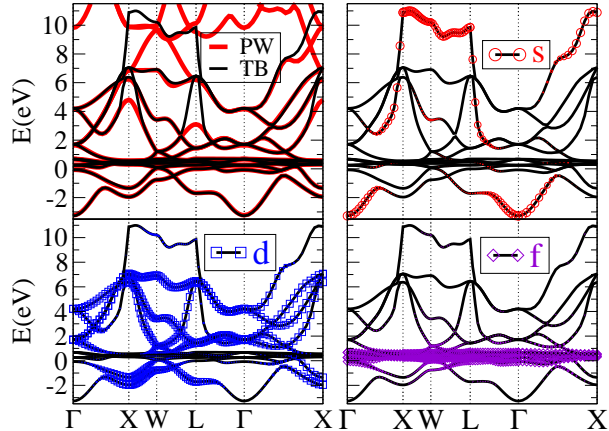}%
\caption{(Color online) Upper left panel shows the comparison between
\textit{ab initio} plane-wave based band structure (red thick lines) and the
downfolded tight-binding result (black thin lines). Other three panels show
projected weight of $6s$, $5d$ and $4f$-orbitals for each band. }%
\label{bd}%
\end{center}
\end{figure}

The bare band Hamiltonian $\mathcal{H}_{0}$ of $\gamma$-Ce is obtained by
downfolding the DFT-LDA band structure from the converged large basis set to a
minimal basis-set ($6s$, $5d$ and $4f$) tight-binding representation with the
recently developed QUasi-Atomic Minimal Basis-set Orbitals (QUAMBOs) scheme
\cite{QUAMBO1,QUAMBO2,QUAMBO3,QUAMBO4}. Fig.\ref{bd} shows that up to 2eV
above Fermi level the tight binding band structure agrees very well with the
LDA result from VASP\cite{VASP} calculations. The orbital projected weight for
each band has also been plotted, confirming that $4f$-contributions are
dominant near the Fermi level. The total number of $4f$ electrons is found to
be 0.93, which is fairly close to one.

\begin{figure}
[ptb]
\begin{center}
\includegraphics[
height=2.1715in,
width=3.2102in
]%
{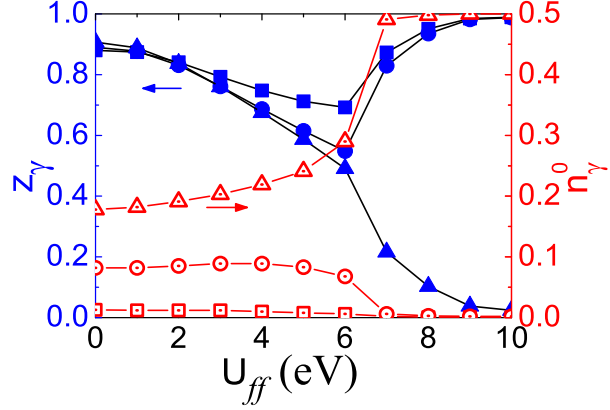}%
\caption{(Color online) Orbital renormalization factors $z_{\gamma}$ (left
axis, blue solid symbols) and orbital occupations $n_{\gamma}^{0}$\ (right
axis, red open symbols) as functions of Coulomb repulsion parameter between
$4f$-electrons $U_{ff}$. Square and circle are for the first two groups of
$4f$-orbitals with degeneracy of 3. The third singly degenerate group is
symbolized by up triangle. For paramagnetic solution we have $z_{\gamma
}=z_{\gamma\uparrow}=z_{\gamma\downarrow}$ and $n_{\gamma}^{0}=n_{\gamma
\uparrow}^{0}=n_{\gamma\downarrow}^{0}$.}%
\label{znf}%
\end{center}
\end{figure}

In order to single out the many-body screening effects of the $5d$-channels,
we first study the system with $\mathcal{H}_{2}$ treated at the Hartree-Fock
mean-field level. The double counting term in Eq.\ref{dc} represents the
Hartree-Fock mean field interactions between the $4f$ and $5d$-electrons
contained in $\mathcal{H}_{0}$, therefore, it exactly cancels the contribution
from the $\mathcal{H}_{2}$\ at the Hartree-Fock level and the whole electron
Hamiltonian of the system reduces to the usual form with many-body Coulomb
interactions only between the $4f$-electrons (i.e., $\mathcal{H}_{1}$).
Proceeding with the Gutzwiller solution of $\mathcal{H}_{1}$, we include in
our calculations $4f$-shell occupations of $f^{0}$, $f^{1}$ and $f^{2}$,
resulting in 106 local $f$-configurations. Fig.\ref{znf} shows the variation
of orbital renormalization factors and orbital occupations with increasing
Coulomb repulsion parameter $U_{ff}$. Due to cubic symmetry of $\gamma$-Ce,
the $4f$-orbitals split into three groups with degeneracy of 3, 3, and 1. It
can be seen that all the $z$-factors initially decrease with increasing
$U_{ff}$\.{. }A transition occurs near $U_{ff}^{T}=6eV$, beyond which the
$z$-factors of the two sets of 3-fold degenerate $4f$-orbitals increase
sharply while the $z$-factor of the singly degenerate $4f$-orbital decreases
rapidly. Meanwhile, the orbital occupations exhibit similar variations near
$U_{ff}^{T}=6eV$. The occupations of the triply degenerate orbital groups
vanish swiftly as $U_{ff}$ exceeds $6eV$, while the singly degenerate orbital
quickly approaches half-filling. These results support the physical picture of
orbital-selective Mott transition\cite{Anisimov02,Koga04} in which the
$f$-electron in the system is redistributed among the $4f$-orbitals such that
the singly degenerate $4f$-orbital approaches Mott localization with a
half-filled band, while the remaining $4f$-orbitals become empty due to strong
$f$-electron correlation effect. However, it has also been shown that a finite
hybridization between the narrow correlated $f$-bands and the wide conduction
bands (mainly 5d bands in this system) can suppress the Mott transition by
Kondo screening\cite{Kotliar05}. The screened $U_{ff}^{scr}$ for $\gamma$-Ce
has been calculated using constrained-LDA method and yields a value of $6eV$
which interestingly coincides with $U_{ff}^{T}$\cite{Macmahan98}. Therefore we
choose the transition point in the $z$-$U$ curve as a fingerprint to identify
the correct input $U_{ff}$\ parameter for $\gamma$-Ce in the following
calculations of different levels.\ We should emphasize that LDA itself can not
capture the many-body Kondo screening effect. The constrained-LDA method gives
reasonable estimation of the screened $U_{ff}^{scr}$ due to other processes,
e.g., dielectric screening.%

\begin{figure}
[ptb]
\begin{center}
\includegraphics[
height=3.9473cm,
width=8.1012cm
]%
{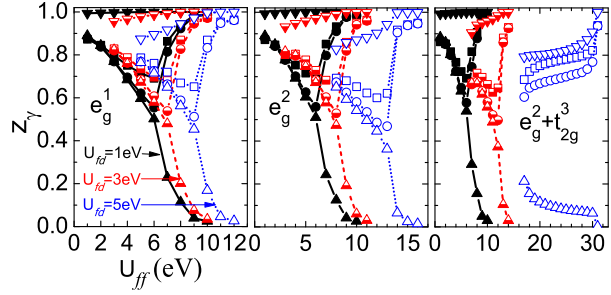}%
\caption{(Color online) The dependence of the orbital renormalization factor
$z_{\gamma}$\ on the Coulomb repulsion parameter between the $4f$-electrons
$U_{ff}$, that between the $4f$ and $5d$-electrons $U_{fd}$, and the number of
many-body screening $5d$-channels explicitly included in the Hamiltonian,
which is 1 for the left panel (one of the $e_{g}$-orbitals), 2 for the middle
(two $e_{g}$-orbitals), and 5 for the right (two $e_{g}$-orbitals and three
$t_{2g}$-orbitals). The set of values of $U_{fd}$\ scanned in the calculation
are 1eV indicated by black solid symbols with solid lines, 3eV by red
half-filled symbols with short-dashes, and 5eV by blue open symbols with short
dots. The three distinct $z$-factors of $4f$-orbitals are depicted by the same
symbols as in Fig.\ref{znf} with the average $z$-factor for the participating
d-channels by down triangles. }%
\label{zfd}%
\end{center}
\end{figure}

\begin{figure}
[ptb]
\begin{center}
\includegraphics[
height=1.5575in,
width=3.1981in
]%
{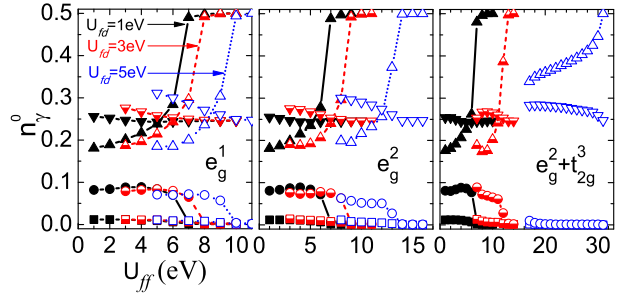}%
\caption{(Color online) The dependence of the orbital occupation $n_{\gamma
}^{0}$\ on the Coulomb repulsion parameter between the $4f$-electrons $U_{ff}%
$, that between the $4f$ and $5d$-electrons $U_{fd}$, and the number of
many-body screening $5d$-channels explicitly included in the Hamiltonian. Plot
settings are same as Fig.\ref{zfd}.}%
\label{nfd}%
\end{center}
\end{figure}

We now study the effect of the many-body screening when the $5d$-electron
screening is explicitly included by treating the $\mathcal{H}_{2}$ and
$\mathcal{H}_{1}$\ on equal footings with Gutzwiller variational wave
functions constructed on the Gutzwiller local subspace consisting of both $4f$
and $5d$-orbitals. Any deviation of the calculation results from the mean
field calculation described above will reflect the effect of many-body
correlation effects arising from including $5d$-electrons in the Gutzwiller
projector. Since we do not expect any significant correlation effect for
$5d$-orbitals, we include all the local $5d$-configurations into account,
which pushes the dimension of the local subspace to 108544. By taking
advantage of sparse matrix techniques, a typical self-consistent calculation
can still be done quite fast on a single processor. A set of Coulomb parameter
$U_{fd}$\ values of \{1eV, 3eV, 5eV\} are chosen with the successive inclusion
of different number of many-body screening $5d$-channels to investigate the
details of the many-body screening process. Our calculation results are shown
in Fig.\ref{zfd} for the orbital renormalization factors $z_{\gamma}$ and in
Fig.\ref{nfd} for the orbital occupations $n_{\gamma}^{0}$. One can see that
the curves shift towards larger values of $U_{ff}$ in a nonlinear fashion with
increasing Coulomb parameter $U_{fd}$ and increasing number of many-body
$5d$-screening channels, indicating that the many-body screening is not a
linear process. The correct input $U_{ff}$ parameter for $\gamma$-Ce in the
calculations which corresponds to the transition point of the $z$-$U$ curve is
pushed much closer to the bare Coulomb repulsion energy between $4f$-electrons
($\sim$23eV based on the neutral atomic $4f$-orbitals) with $U_{fd}\approx5eV$
and all the $5d$-channels participating in many-body screening process. It
should be pointed out that we encounter some numerical problem in searching
solutions of the system on the smaller $U_{ff}$ side, however, this numerical
difficulty will not affect our conclusion.

\begin{figure}
[ptb]
\begin{center}
\includegraphics[
height=6.8688cm,
width=8.099cm
]%
{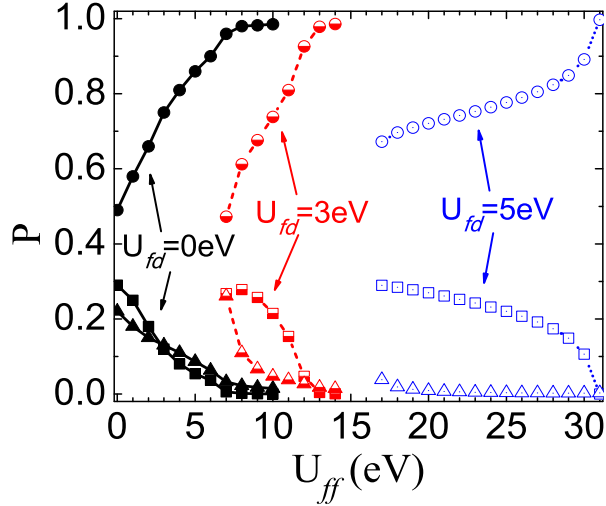}%
\caption{(Color online) The sum of occupation probabilities of some group of
local electronic configurations as a function of $U_{ff}$ and $U_{fd}$ with
all the $5d$ and $4f$-orbitals in the Gutzwiller local subspace. Squares are
for the configurations with no $f$-electron occupied, circles for those with
one $f$-electron, and up triangles for those with two $f$-electrons. $U_{fd}%
$=0eV is indicated by black solid symbols with solid lines, 3eV by red
half-filled symbols with short-dashes, and 5eV by blue open symbols with short
dots. The results with $U_{fd}$=0 are same as those with only $4f$ orbitals
included in the Gutzwiller local subspace since there are no Coulomb
interactions between $5d$ and $4f$-electrons.}%
\label{pf012}%
\end{center}
\end{figure}

To shed some light on the many-body screening mechanism, it is instructive to
examine the variations of the local electronic configuration occupation
probabilities $\left\{  p_{\mathcal{F}}\right\}  $, which are the new
variational degrees of freedom in the Gutzwiller treatment. However, a
straightforward analysis of the local electronic configurations can be rather
tedious since its dimension is usually very large. In Fig.\ref{pf012} we
present the sum of the occupation probabilities of the grouped local
configurations according to the number of $f$-electrons occupied, which is
closely related to the $z$-factors for the $4f$-orbitals. The local
configurations with two $f$-electrons occupied can survive to larger $U_{ff}$
and the local configurations with one $f$-electron occupied will approach a
full occupation at higher $U_{ff}$ by increasing $U_{fd}$. This tendency is
consistent with the behavior of the $z$-factors for the $4f$-electrons.
Another way to look at the many-body screening mechanism is to investigate the
response of the screening $5d$-electrons when the many-body screening is
turned on. Fig.\ref{zfd} shows that the $z$-factors for $5d$-orbitals are
smaller than one, but are still quite large ($\gtrsim0.8$). It implies that
the $5d$-electrons rearrange themselves in the local subspace in a way
deviating from Hartree-Fock mean field solution in order to effectively screen
the interactions between the $4f$-electrons.

By including Ce $5d$ and $4f$-orbitals into the Gutzwiller local subspace for
the construction of Gutzwiller wave function., i.e., having $5d$-electrons
treated in the same Gutzwiller level as $4f$-electrons beyond static
Hartree-Fock mean-field approximation, we observe that the equivalent $U_{ff}%
$\ is pushed towards its bare atomic value in $\gamma$-Ce. Therefore local
many body response in the $5d$-channels is the dominant contribution in the
screening process for $4f$-electrons in bulk Ce. This justifies our method in
using single-site Gutzwiller treatment in this system. In some cases where
significant screening effect comes from the atomic environment\cite{Brink95},
we expect the generalized cluster-Gutzwiller treatment in parallel with
cluster-DMFT would be useful\cite{Maier05}.

\section{Conclusion}

A method for explicitly including many-body screening process into
self-consistent calculations based on Gutzwiller variational wave function and
Gutzwiller approximation has been proposed and applied to the tight-binding
electronic structure calculation of $\gamma$-Ce. The resultant Coulomb
repulsion parameter between $4f$-electrons $U_{ff}$\ can be much larger than
the corresponding screened $U_{ff}$ with moderate value of Coulomb repulsion
parameter between $4f$ and $5d$-electrons $U_{fd}$. Assuming the dominant
many-body screening effect for the strongly correlated orbitals (e.g.,
$4f$-orbitals for Ce) can be explicitly taken into account in this local
approach, it is therefore of great interest to examine how the current scheme
performs in full DFT-based self-consistent calculations where all the Coulomb
repulsion energies between various orbitals in the Gutzwiller local subspace
are directly obtained from atomic orbital integrations. Since the atomic
orbitals are generally non-orthogonal, it is also highly worthwhile to extend
the scheme from orthogonal orbitals to non-orthogonal orbitals. The proposed
scheme here furnishes a promising way towards the realization of a
self-consistent \textit{ab initio} GDFT.

\begin{acknowledgments}
We are grateful to J. Schmalian for useful discussions. Work at the Ames
laboratory was supported by the U.S. Department of Energy, Office of Basic
Energy Science, Division of Materials Science and Engineering including a
grant of computer time at the National Energy Research Supercomputing Center
(NERSC) at the Lawrence Berkeley National Laboratory under Contract No. DE-AC02-07CH11358.
\end{acknowledgments}

\end{document}